\documentclass[letterpaper,english,reprint,nofootinbib,superscriptaddress]{revtex4-1}
\usepackage[T1]{fontenc}
\usepackage[latin9]{inputenc}
\usepackage{color}
\usepackage{float}
\usepackage{amsmath}
\usepackage{amssymb}
\usepackage{graphicx,psfrag}
\usepackage{ulem}


\usepackage{babel}
\begin{document}
\title{Neutron Stars in Modified Teleparallel Gravity}
\author{S. G. Vilhena}
\email{vilhenasg@gmail.com}
\affiliation{Departamento de F\'isica, Instituto Tecnol\'ogico de Aeron\'autica Pra\c{c}a
Mal. Eduardo Gomes 50 CEP 12228-900 S\~ao Jos\'e dos Campos, S\~ao Paulo,
Brazil}
\affiliation{Centro Brasileiro de Pesquisas F\'isicas, Rua Dr. Xavier Sigaud, 150
URCA, Rio de Janeiro CEP 22290-180, RJ, Brazil}

\author{S. B. Duarte}
\email{sbd@cbpf.br}

\affiliation{Centro Brasileiro de Pesquisas F\'isicas, Rua Dr. Xavier Sigaud, 150
URCA, Rio de Janeiro CEP 22290-180, RJ, Brazil}

\author{M. Dutra}
\email{marianad@ita.br}

\affiliation{Departamento de F\'isica, Instituto Tecnol\'ogico de Aeron\'autica Pra\c{c}a
Mal. Eduardo Gomes 50 CEP 12228-900 S\~ao Jos\'e dos Campos, S\~ao Paulo,
Brazil}
\affiliation{Univ Lyon, Univ Claude Bernard Lyon 1, CNRS/IN2P3, IP2I Lyon, UMR 5822, F-69622, Villeurbanne, France}

\author{P. J. Pompeia}
\email{pompeia@ita.br}

\affiliation{Departamento de F\'isica, Instituto Tecnol\'ogico de Aeron\'autica Pra\c{c}a
Mal. Eduardo Gomes 50 CEP 12228-900 S\~ao Jos\'e dos Campos, S\~ao Paulo,
Brazil}

\date{\today}

\begin{abstract}
We investigate compact objects in modified teleparallel gravity with realistic equations of state. We propose a modification of Teleparallel Equivalent to General Relativity, then an appropriate tetrad is applied to the field equations. A specific set of relations showing an equivalency between our gravitational model and the New General Relativity is found. The conservation equation implies that our Tolman-Oppenheimer-Volkoff equations are presented with an effective gravitational coupling constant. Numerical analysis using realistic equations of state is made, and the behavior of mass, radius, and the relation mass-radius as functions of a free parameter of our model is also investigated.
\end{abstract}

\maketitle

\section{\label{sec:level1}Introduction\protect \\
}

Although General Relativity (GR) is successful in describing gravitational phenomena, it may not be the final theory of gravity. Some unanswered questions still remain, like the $H_{0}$ tension~\cite{Valentino2021,Efstathiou2021,Valentino2017,Verde2019,Vagnozzi2020} and the $\sigma_{8}$ problem~\cite{Valentino2017,Verde2019,Nunes2021}. Some of the failures of the standard GR (in particular, the accelerated expansion of the Universe and galaxy rotation curves) may be solved by the proposition of the existence of exotic matter (dark matter) and dark energy, both of them composing the so-called ``dark sector''~\cite{Valentino2021,Valentino2017,Efstathiou2021,Verde2019,Nunes2021,Slozar2019,Huterer2017,Leo(2020)}. However, even in the presence of the dark sector, some problems still
remain in high-energy regimes, where quantization is expected  \citep{Stelle1977,Anselmi(2007)}. 

Modified theories of gravity have long been studied in the search of alternative frameworks to the dark sector~\citep{Langlois(2018),Kase(2019),Finch(2018),Motta(2021),Rahaman(2014),Zubair2022a}, as well as in the search for a quantizable theory. One way of modifying gravity consists of replacing the original Einstein-Hilbert Lagrangian by an arbitrary function of its original argument, as has been done, for instance, in $f\left(R\right)$ theories \citep{Sotiriou2010,DeFelice2010,Nojiri2011,Naf2011,DeLaurentis2013,Berry2011,Gottlober1991,Astashenok2021a,Astashenok2020,Astashenok2021b}. An example is Starobinsky's model, where he proposed a modification of the Einstein-Hilbert action by including a square term $R^{2}$, which could explain the early inflation of the universe~\citep{Starobinsky(1979),Starobinsky1979,Gurovich1979,Oikonomou2021}. As can be verified in several works, these modified theories are usually good candidates to replace the dark energy sector~\citep{Motta(2021),Rahaman(2014)}. Sometimes these theories may fail in solar system scales, but in several cases, they are still useful in the context of effective theories. Applications of modified theories of gravity have been discussed in several areas like compact objects, the early universe, gravitational waves, and so on~\citep{Leo(2020),Capozziello2011,PRD2021,Starobinsky(1979),Starobinsky1979,Dutra(2020),Zhai(2022),Pradhan(2022),Astashenok(2021),Lobato(2021),Duarte(2022a),Duarte(2022b),Panah2016,Panah2017,Panah2019,Panah2022,Zubair2021a}.

In the Riemannian manifold, where GR is constructed, several proposals modify the action by the inclusion of invariants involving the curvature (e.g. $R_{\mu\nu}R^{\mu\nu}$, $C_{\mu\nu\rho\sigma}C^{\mu\nu\rho\sigma}$) and eventually its derivatives~\citep{Sotiriou2010,Gottlober1991,PRD2016,PRD2019,Quarteto(2019)} (e.g. $R\Box R,\nabla_{\mu}R\nabla^{\mu}R$). If instead of using a Riemann manifold, another one is chosen, then other theories of gravity are obtained. These theories may or may not have an equivalent
formulation in a Riemannian manifold. An example is the Teleparallel Equivalent to General Relativity (TEGR), a theory built in the Weitzenb\"ock manifold which proves to be equivalent theory to GR. In other words, the same predictions and results obtained in GR are also obtained in TEGR~\citep{Aldrovandi(1995),Andrade(2000),Bahamonde(2021)}.

Whilst in Riemann manifolds, the spacetime connection is completely determined by the Christoffel symbols (which in their turn are dictated by the metric tensor), in Weitzenb\"ock spaces, two connections appear: One is the Weitzenb\"ock connection, related to the gauge symmetry associated with the translation group; the other is the spin connection, associated with local Lorentz transformations~\citep{Krrsak(2019),Cai(2016),Aldrovandi(1995),Andrade(2000),Bahamonde(2021)}. When the absolute parallelism condition is adopted, the spin connection vanishes, and the Weitzenb\"ock connection is determined by the tetrad, which plays the role of fundamental field.
In this case, however, there is a price to be paid: The loss of local Lorentz invariance may restrict the equivalence of solutions found for different tetrad fields. These problems are avoided when ``good tetrads'' are used~\citep{Tamanini(2012),Krssak(2016),Tamanini(2013)}.
As a consequence, while in Riemann manifolds gravity is manifest by curvature, in Weitzenb\"ock manifolds the spacetime is characterized by the torsion tensor. In TEGR, for instance, the Lagrangian is essentially the scalar torsion, $T$, a quadratic combination of the torsion tensor
obtained by the contraction of the torsion tensor $T_{\rho\mu\nu}$ with $S^{\rho\mu\nu}$, which is built as a specific linear combination of $T_{\rho\mu\nu}$ that allows us to recover the results of GR~\citep{Cai(2016),Aldrovandi(1995),Andrade(2000),Bahamonde(2021),Tamanini(2012),Krssak(2016),Tamanini(2013)}.

In this perspective, this work proposes a modification of  the integral action of TEGR action by the introduction of free parameters on the three quadratic invariants that compose $T$. Essentially, we replace $S^{\rho\mu\nu}$ by $\Sigma^{\rho\mu\nu}$, a general linear combination
of $T_{\rho\mu\nu}$ and its trace. This model can be properly described in the context of New General Relativity (NGR)~\citep{Bahamonde(2021),Hayashi(1979)}, where different representations of the quadratic invariants are considered.

To test the validity of the proposed model, a physical system has to be considered. A good ``laboratory'' to test different theories of gravity can be found in compact objects. Besides allowing us to test different gravitational models, they also permit us to study fundamental properties of matter. Compact objects are usually studied by the Tolman-Oppenheimer-Volkoff (TOV) equations~\cite{tov1,tov2} (or their generalization), which connect the dynamics of the gravitational field with the energy content of the object, the latter strongly dependent on the equation of state (EoS) of the type of matter considered~\citep{Dutra(2020),Zhai(2022),Pradhan(2022),Astashenok(2021),Lobato(2021),Duarte(2022a),Duarte(2022b),Araujo(2021),Araujo(2021a)}. For the zero temperature regime, detailed knowledge of the equations of state of hadronic models, both relativistic and non-relativistic, becomes fundamental in the description, for example, of neutron stars (NS), studied at densities above six times the saturation density of nuclear matter~\cite{Sci304-536-2004}. Properties of these objects, such as the mass-radius relation, are directly influenced by particular features of each hadronic model used~\cite{PRC93-025806(2016)}. It is important to mention that an important source of information about the characteristics of astrophysical systems, such as NS, is NASA's Neutron star Interior Composition Explorer (NICER)~\cite{nicer}. From the data extracted from this experiment, such as PSR~J0030+0451~\cite{Miller(2019),Riley(2019)} and PSR~J0740+6620~\cite{Miller(2021),Riley(2021)} it is possible to make estimates about the mass-radius profile of NS. Additionally, the data coming from gravitational wave detections are extremely relevant, such as the LIGO-VIRGO (LVC)~\cite{era} with the data coming from the GW170817~\cite{era,Abbott(2017),Abbott(2018),Cowperthwaite(2017)} and more recently GW190814~\cite{Abbott(2020)}, and GW190425~\cite{Abbott(2020)-2}.

Here, we will use the nonlinear Walecka model~\cite{AnnPhy83-491-1974,Ring1980,NPA292-413-1977,NPA656-331-1999} in the mean-field approximation, one of the main representatives of the relativistic hadronic models to describe neutron stars. In its simplest version, this model considers protons and neutrons as fundamental particles interacting with each other through the exchange of the scalar meson $\sigma$ and vector $\omega$, which physically represent the attractive and repulsive part, respectively, of the nuclear interaction. In this model, the free parameters present in the theory are adjusted to reproduce, at zero temperature, the quantities obtained by many-body physics, such as the binding energy, incompressibility, and saturation density of infinite nuclear matter~\cite{PRC90-055203-2014}.

This paper is structured as follows: In Sec.~\ref{sec:Modified-Teleparallel-Gravity}, we present the modification of the teleparallel Lagrangian and the field equations obtained. A constraint on the free parameters is obtained to preserve the conservation equation. In the last part of the section, we demonstrate that our model and NGR are equivalent under a specific
relation of the parameters from both models. The gravitational theory provides a set of equations \textendash{} TOV equations \textendash{} that can be used to model the structure of compact objects. However, this set of equations is incomplete, hence it is necessary to specify
an EoS for this system. Once the EoS is chosen, the next step is to solve this set of equations. In our model, this system is solved numerically using realistic EoS, including the free parameter $\beta_3$. In Sec.~\ref{rmf}, we will present the description of the relativistic mean-field model (RMF) used in this work. In Sec.~\ref{sec:Numerical-Analysis}, numerical analysis is performed and the behavior of the mass and the radius of the object
as functions of the free parameter $\beta_3$ are determined. In the same
section, observational data are employed in order to analyze how our model behaves under the two EoSs. Our final comments are presented in Sec.
\ref{sec:Final-Remarks}.

\section{Modified Teleparallel Gravity and Field Equations\label{sec:Modified-Teleparallel-Gravity}}

The TEGR action is given by

\begin{equation}
S=-\frac{1}{2\chi}\int eS^{\rho\mu\nu}T_{\rho\mu\nu}d^{4}x+S_{m}\,,\label{eq:Act_TEGR}
\end{equation}
where $S_{m}$ stands for an action for matter fields and $e=\det e_{\hphantom{a}\mu}^{a}$, $\chi=8\pi$, with $G=\hbar=c=1$. The tensor $S^{\mu\nu\rho}$ is defined as
\begin{equation}
S^{\mu\nu\rho}\equiv\frac{1}{4}\left(T^{\mu\nu\rho}+T^{\nu\mu\rho}-T^{\rho\mu\nu}\right)+\frac{1}{2}\left(g^{\mu\rho}T^{\nu}-g^{\mu\nu}T^{\rho}\right)\,,\label{eq:Sigma}
\end{equation}
 where $T^{\rho\mu\nu}$ is the torsion built with the Weitzenb\"ock connection,
 \begin{equation}
 T_{\hphantom{\alpha}\mu\nu}^{\alpha}\equiv\Gamma_{\mu\nu}^{\alpha}-\Gamma_{\nu\mu}^{\alpha}=e_{a}^{\alpha}\left(\partial_{\mu}e_{\nu}^{a}-\partial_{\nu}e_{\mu}^{a}\right)
 \end{equation}
and $T_{\hphantom{\mu}\mu}^{\mu\hphantom{\mu}\rho}\equiv T^{\rho}$ is its trace.

Varying the action (\ref{eq:Act_TEGR}) with respect to $e_{\hphantom{a}\mu}^{a}$
leads us to the field equations
\begin{equation}
\partial_{\rho}\left(4eS_{f}^{\hphantom{f}\lambda\rho}\right)+4eS_{d}^{\hphantom{d}\lambda\rho}T_{\hphantom{d}f\rho}^{d}-ee_{f}^{\lambda}S_{\mu\nu\rho}T^{\mu\nu\rho}=-2\chi ee_{f}^{\rho}T_{\hphantom{\lambda}\rho}^{\lambda}\,,\label{eq:FieldEq}
\end{equation}
where $T_{\hphantom{\lambda}\rho}^{\lambda}$ is the energy-momentum tensor. The action (\ref{eq:Act_TEGR}) can be modified as has been done, for instance, in Ref.~\citep{Hayashi(1979)}, where the authors proposed a modification of the Lagrangian with a quadratic combination of irreducible objects of the torsion decomposition. In this work, we modify the teleparallel Lagrangian by introducing parameters in the torsion scalar such that the new action reads 
\begin{eqnarray}
S&=-\frac{1}{2\chi}\int e\left(\beta_3T^{\rho}T_{\rho}-\frac{\beta_1}{4}T^{\rho\mu\nu}T_{\rho\mu\nu}-\frac{\beta_2}{2}T^{\rho\mu\nu}T_{\mu\rho\nu}\right)d^{4} x \nonumber \\
&+S_{m}\,,\label{eq:GeneLag}
\end{eqnarray}
In Eq. (\ref{eq:GeneLag}) it is clear that we obtain Eq. (\ref{eq:Act_TEGR}) when we input the values $\beta_1=\beta_2=\beta_3=1$. In this modified teleparallel gravity, we obtain the same structure of the field equations in Eq. (\ref{eq:FieldEq}) when we modify Eq. (\ref{eq:Sigma}) to
\begin{eqnarray}
&S^{\mu\nu\rho}\rightarrow\Sigma^{\mu\nu\rho}\,,\nonumber\\
&\Sigma^{\mu\nu\rho}\equiv\frac{\beta_1}{4}T^{\mu\nu\rho}+\frac{\beta_2}{4}\left(T^{\nu\mu\rho}-T^{\rho\mu\nu}\right)\nonumber\\
&+\frac{\beta_3}{2}\left(g^{\mu\rho}T^{\nu}-g^{\mu\nu}T^{\rho}\right)\,,   
\label{eq:NewSigma}
\end{eqnarray}
so that the parameters $\beta_1,\beta_2,\beta_3$ are hidden parameters, not appearing explicitly in the field equations:
\begin{equation}
\partial_{\rho}\left(4e\Sigma_{f}^{\hphantom{f}\lambda\rho}\right)+4e\Sigma_{d}^{\hphantom{d}\lambda\rho}T_{\hphantom{d}f\rho}^{d}-ee_{f}^{\lambda}\Sigma_{\mu\nu\rho}T^{\mu\nu\rho}=-2\chi ee_{f}^{\rho}T_{\hphantom{\lambda}\rho}^{\lambda}\,.\label{eq:NewFieldEq}
\end{equation}

Our intention is to construct a model of static compact objects, hence, considering a static and spherically symmetric line element $ds^{2}=\gamma_{00}^{2}dt^{2}-\gamma_{11}^{2}dr^{2}-r^{2}d\theta^{2}-r^{2}\sin^{2}\theta d\varphi^{2}\,.$
Thus, the vierbein or tetrad in the Schwarzschild coordinate systems is given by \cite{Zhai(2022)}

\begin{equation}
e_{\hphantom{a}\mu}^{a}=\left(\begin{array}{cccc}
\gamma_{00} & 0 & 0 & 0\\
0 & \gamma_{11}\sin\theta\cos\phi & r\cos\theta\cos\phi & -r\sin\theta\sin\phi\\
0 & \gamma_{11}\sin\theta\sin\phi & r\cos\theta\sin\phi & r\sin\theta\cos\phi\\
0 & \gamma_{11}\cos\theta & -r\sin\theta & 0
\end{array}\right)\,.\label{eq:Tetamu}
\end{equation}

The tetrad in Eq. (\ref{eq:Tetamu}) is non-diagonal, but it is considered a ``good tetrad'' (see Ref. \citep{Tamanini(2012)} for a discussion about ``good'' and ``bad'' tetrads). Using the tetrad above in Eq. (\ref{eq:NewFieldEq}), and assuming $T_{\hphantom{1}1}^{1}=T_{\hphantom{2}2}^{2}=T_{\hphantom{3}3}^{3}=-p$,
$T_{\hphantom{0}0}^{0}=\rho$, it is possible to demonstrate that the sixteen equations are reduced to only three, namely,
\begin{equation}
\begin{cases}
\chi\gamma_{11}^{3}\rho r^{2}=2\beta_3\gamma_{11}^{\prime}r+\beta_3\gamma_{11}^{3}-\frac{1}{2}\left(\beta_2+\beta_1\right)\gamma_{11}\\
\qquad+\left(2\beta_3-\beta_2-\beta_1\right)\left[\left(\frac{\gamma_{00}^{\prime}\gamma_{11}^{\prime}}{\gamma_{00}}+\frac{\gamma_{00}^{\prime\prime}\gamma_{11}}{4\gamma_{00}^{2}}-\frac{\gamma_{00}^{\prime\prime}\gamma_{11}}{2\gamma_{00}}\right)r^{2}\right.\\
\qquad\qquad\qquad\left.-\frac{\gamma_{00}^{\prime}\gamma_{11}}{\gamma_{00}}r-\gamma_{11}^{2}+\gamma_{11}^{3}\right]\\
-\chi\gamma_{00}\gamma_{11}^{2}pr^{2}=\beta_3\left(2\gamma_{00}^{\prime}r-\gamma_{00}\gamma_{11}^{2}+\gamma_{00}\right)\\
\qquad+\left(2\beta_3-\beta_2-\beta_1\right)\left(\gamma_{00}^{\prime2}r^{2}-\frac{1}{2}\gamma_{00}\gamma_{11}^{2}+\frac{1}{2}\gamma_{00}\right)\\
-\chi\gamma_{00}\gamma_{11}^{3}pr=\beta_3\left(\gamma_{00}^{\prime}\gamma_{11}^{\prime}r-\gamma_{00}^{\prime\prime}\gamma_{11}r+\gamma_{00}\gamma_{11}^{\prime}-\gamma_{00}^{\prime}\gamma_{11}\right)\\
\qquad+\frac{1}{2}\left(2\beta_3-\beta_2-\beta_1\right)\times \\ \qquad\quad \left(\gamma_{00}^{\prime}\gamma_{11}^{2}+\frac{\gamma_{00}^{\prime2}\gamma_{11}}{2\gamma_{00}}r+\gamma_{00}\gamma_{11}^{\prime}-\gamma_{00}^{\prime}\gamma_{11}\right)
\end{cases}\,.\label{eq:equationsinde}
\end{equation}
We still have to recall that the conservation equation is valid, and reads
\begin{equation}
p^{\prime}=-\left(p+\rho\right)\frac{1}{\gamma_{00}}\gamma_{00}^{\prime}\,.\label{eq:Conserlaw}
\end{equation}
In TEGR, the equations equivalent to Eq~(\ref{eq:equationsinde}) and Eq.~(\ref{eq:Conserlaw}) are not a set of independent equations. There, a standard procedure consists in manipulating three of the equations and showing that the fourth is obtained. Here, we also have four equations that are not independent; if we manipulate the field equations in Eq.~(\ref{eq:equationsinde}), analogously to what is made in TEGR, we obtain:
\begin{align}
 & \frac{1}{2}\chi\gamma_{11}^{3}r^{2}\left[\gamma_{00}p^{\prime}+\gamma_{00}^{\prime}\left(p+\rho\right)\right]=\nonumber \\
 & \left(2\beta_3-\beta_2-\beta_1\right)\left[\gamma_{00}^{\prime2}\gamma_{11}r+\gamma_{00}^{\prime\prime}\gamma_{00}^{\prime}\gamma_{11}r^{2}+\frac{\gamma_{00}^{\prime2}\gamma_{11}}{4\gamma_{00}}r\right.\nonumber \\
 & -\gamma_{00}^{\prime2}\gamma_{11}^{\prime}r^{2}+\left.\frac{1}{2}\frac{\gamma_{00}^{\prime2}}{\gamma_{00}}\gamma_{11}^{\prime}r^{2}+\frac{\gamma_{00}^{\prime\prime}}{8\gamma_{00}^{2}}\gamma_{00}^{\prime}\gamma_{11}r^{2}-\frac{\gamma_{00}^{\prime2}\gamma_{11}}{2\gamma_{00}}r\right.\nonumber\\
 & \left.-\frac{\gamma_{00}^{\prime\prime}}{4\gamma_{00}}\gamma_{00}^{\prime}\gamma_{11}r^{2}\right]\,.\label{eq:conservationseq}
\end{align}
Eq.~(\ref{eq:conservationseq}) shows that the conservation equation continues to be valid only if the right-hand side is null. In other words, our model has to obey the following constraint: 
\begin{equation}
2\beta_3-\beta_2-\beta_1=0\,.\label{eq:constraint}
\end{equation}
This constraint is not only necessary but is very useful since we can rewrite the equations in Eq. (\ref{eq:equationsinde}) in a way that they depend only on one of the free parameters, namely, $\beta_3$. We cannot state that the relation in Eq.~(\ref{eq:constraint}) is valid in general, but it is valid in particular systems with symmetries as we see in the present case. We obtain a simplified set of equations
\begin{equation}
\begin{cases}
\bar{\chi}\gamma_{11}^{3}\rho r^{2}=2\gamma_{11}^{\prime}r+\gamma_{11}^{3}-\gamma_{11}\\
-\bar{\chi}\gamma_{00}\gamma_{11}^{2}pr^{2}=2\gamma_{00}^{\prime}r-\gamma_{00}\gamma_{11}^{2}+\gamma_{00}\\
-\bar{\chi}\gamma_{00}\gamma_{11}^{3}pr=\gamma_{00}^{\prime}\gamma_{11}^{\prime}r-\gamma_{00}^{\prime\prime}\gamma_{11}r+\gamma_{00}\gamma_{11}^{\prime}-\gamma_{00}^{\prime}\gamma_{11}
\end{cases}\,,\label{eq:ConjC}
\end{equation}
where we have defined an effective gravitational coupling constant as
\begin{equation}
\bar{\chi}=\frac{\chi}{\beta_3}\,.\label{eq:rhopeffec}
\end{equation}
The role of the parameter $\beta_3$ is to modify the intensity of the gravitational coupling between matter and the gravitational field. When taking into account the conservation equation, we can work with another set of equations, which equivalently represents the same system. We choose to work with the following equations:
\begin{equation}
\begin{cases}
\bar{\chi}\gamma_{11}^{3}\rho r^{2}=2\gamma_{11}^{\prime}r+\gamma_{11}^{3}-\gamma_{11}\\
-\bar{\chi}\gamma_{00}\gamma_{11}^{2}pr^{2}=2\gamma_{00}^{\prime}r-\gamma_{00}\gamma_{11}^{2}+\gamma_{00}\\
p^{\prime}=-\left(p+\rho\right)\frac{1}{\gamma_{00}}\gamma_{00}^{\prime}
\end{cases}\,.\label{eq:Conj_bar}
\end{equation}
Note that the last equation is essentially Eq.(\ref{eq:Conserlaw}) divided by $\beta_3$. Now, we have three equations expressed in terms of the effective gravitational constant, the pressure, and the energy density. These equations read exactly like the ones in TEGR, the difference among them being the free parameter $\beta_3$ dividing $\chi$. In a vacuum, we wouldn't see a difference between the equations of our model and those of TEGR or GR -- as consequence, we don't expect modifications of the general solutions in the study of planetary motions, deflection of light rays, and other tests in absence of matter. However, when imposing boundary conditions for these problems (which typically involve the energy-momentum tensor) the discrepancies between these models appear due to the differences in the gravitational coupling. The structure of the equations is such that it allows similar boundary conditions on a star
as those used in GR.\footnote{The authors are grateful to an anonymous referee whose comments allowed us to improve the discussion concerning the role of the effective coupling constant.}

By introducing a change of variable, 
\begin{equation}
\frac{1}{\gamma_{11}^{2}}=1-\frac{2u\left(r\right)}{r}\,,\label{eq:ChaVar}
\end{equation}
the first equation of (\ref{eq:Conj_bar}) reads
\begin{equation}
u^{\prime}=\frac{1}{2}\bar{\chi}\rho r^{2}\,.\label{eq:ubarprime}
\end{equation}
By combining the last two equations of Eq. (\ref{eq:Conj_bar}) with Eq. (\ref{eq:ChaVar}),
we conclude that
\begin{equation}
p^{\prime}=-\frac{p+\rho\left(p\right)}{\left(r^{2}-2ur\right)}\left(\frac{1}{2}\bar{\chi}pr^{3}+u\right)\,.\label{eq:pbarprime}
\end{equation}
As expected, Eqs. (\ref{eq:ubarprime}) and (\ref{eq:pbarprime}) are analogous to those of TEGR with $\chi$ renormalized by the free parameter $\beta_3$. Both equations and an EoS determine completely the mass distribution of a compact object.

Here, the proposed model can also be expressed from an NGR perspective. In that approach, the Lagrangian is constructed using the vector, axial, and tensor decompositions of the torsion:
\begin{equation}
\begin{cases}
\mathfrak{v}_{\mu} & =T_{\hphantom{\nu}\nu\mu}^{\nu}\\
\mathfrak{a}_{\mu} & =\frac{1}{6}\epsilon_{\mu\nu\rho\sigma}T^{\nu\rho\sigma}\\
\mathfrak{t}_{\rho\mu\nu} & =T_{\left(\mu\nu\right)\rho}+\frac{1}{3}\left(T_{\hphantom{\sigma}\sigma}^{\sigma}{}_{\left(\mu\right.}g_{\left.\nu\right)\rho}-T_{\hphantom{\sigma}\sigma\rho}^{\sigma}g_{\mu\nu}\right)
\end{cases}\,.\label{eq:vatdecom}
\end{equation}
In NGR, three scalars are constructed with the decompositions above, namely,
\begin{equation}
T_{vec}=\mathfrak{v}_{\mu}\mathfrak{v}^{\mu},T_{axi}=\mathfrak{a}_{\mu}\mathfrak{a}^{\mu},T_{ten}=\mathfrak{t}_{\lambda\mu\nu}\mathfrak{t}^{\lambda\mu\nu}\,.\label{eq:TEscs}
\end{equation}
A linear combination of these scalars gives us the action of the NGR, it reads
\begin{equation}
S=-\frac{1}{2\chi}\int e\left(v_{vec}T_{vec}+v_{axi}T_{axi}+v_{ten}T_{ten}\right)d^{4}x+S_{m}\,,\label{eq:Act_NGR}
\end{equation}
The action integral Eq. (\ref{eq:Act_NGR}) is equivalent to our proposed model on Eq.~(\ref{eq:GeneLag}), when we identify
\begin{equation}
\begin{cases}
\beta_1=2v_{ten}-\frac{4v_{axi}}{18}\\
\beta_2=v_{ten}+\frac{4v_{axi}}{18}\\
\beta_3=\frac{v_{ten}}{2}-v_{vec}
\end{cases}\,.\label{Conj_Par_NRG_abc}
\end{equation}
As stated above, the relation among our parameters, expressed by Eq.~(\ref{eq:constraint}), is important to keep the conservation equation valid. In terms of the parameters of NGR, the constraint given in Eq. (\ref{eq:constraint}) reads
\begin{equation}
v_{vec}+v_{ten}=0\,.\label{eq:Cond_LC_NRG}
\end{equation}
This result expresses that the axial part of the decomposition of the torsion plays no role to keep the conservation equation valid.

\section{Equation of State\label{rmf}}
In this section, we describe the RMF model that we will use to generate the radius-mass profile of the NS. The Lagrangian density that describes the nonlinear Walecka model~\cite{mdi3,PRC90-055203-2014} taking into account the leptons (electron and muon) is given by
\begin{align}
\mathcal{L} &= \overline{\psi}(i\gamma^\mu\partial_\mu - M_{\mbox{\tiny nuc}})\psi 
+ g_\sigma\sigma\overline{\psi}\psi 
- g_\omega\overline{\psi}\gamma^\mu\omega_\mu\psi
\nonumber \\ 
&- \frac{g_\rho}{2}\overline{\psi}\gamma^\mu\vec{\rho}_\mu\vec{\tau}\psi
+\frac{1}{2}(\partial^\mu \sigma \partial_\mu \sigma - m^2_\sigma\sigma^2)
- \frac{A}{3}\sigma^3 - \frac{B}{4}\sigma^4 
\nonumber\\
&-\frac{1}{4}F^{\mu\nu}F_{\mu\nu} 
+ \frac{1}{2}m^2_\omega\omega_\mu\omega^\mu 
+ \frac{C}{4}(g_\omega^2\omega_\mu\omega^\mu)^2 -\frac{1}{4}\vec{B}^{\mu\nu}\vec{B}_{\mu\nu} 
\nonumber \\
&+ g_\sigma g_\rho^2\sigma\vec{\rho}_\mu\vec{\rho}^\mu
\left(\alpha_2+\frac{1}{2}\alpha'_2g_\sigma\sigma\right) 
+ \frac{1}{2}\alpha'_3g_\omega^2 g_\rho^2\omega_\mu\omega^\mu
\vec{\rho}_\mu\vec{\rho}^\mu
\nonumber\\
&+ \frac{1}{2}m^2_\rho\vec{\rho}_\mu\vec{\rho}^\mu
+ g_\sigma g_\omega^2\sigma\omega_\mu\omega^\mu
\left(\alpha_1+\frac{1}{2}\alpha'_1g_\sigma\sigma\right) 
\nonumber \\
&+ \sum_{l=e,\mu}\overline{\psi_l}(i\gamma^\mu\partial_\mu - m_{\mbox{\tiny l}})\psi_l,
\label{lagrangiana}
\end{align}
where the nucleon rest mass is $M$ and the mesons masses are $m_\sigma$, $m_\omega$, and $m_\rho$. $F_{\mu\nu}=\partial_\nu\omega_\mu-\partial_\mu\omega_\nu$ and $\vec{B}_{\mu\nu}=\partial_\nu\vec{\rho}_\mu-\partial_\mu\vec{\rho}_\nu- g_\rho (\vec{\rho}_\mu \times \vec{\rho}_\nu)$. The last term of the Eq.~\eqref{lagrangiana} represents the leptons part, with $l = e (\mu)$ for the electron (muon). With the Euler-Lagrange equations and the mean-field approximation for the fields, we find the  equations for pressure and energy~\cite{PRC90-055203-2014}, given respectively by 
\begin{align}
&p = - \frac{1}{2}m^2_\sigma\sigma^2 
- \frac{A}{3}\sigma^3 - \frac{B}{4}\sigma^4 + \frac{1}{2}m^2_\omega\omega_0^2 
+ \frac{C}{4}(g_\omega^2\omega_0^2)^2 
\nonumber\\
&+ g_\sigma g_\omega^2\sigma\omega_0^2
\left(\alpha_1 + \frac{\alpha'_1g_\sigma\sigma}{2}\right) 
+ g_\sigma g_\rho^2\sigma\bar{\rho}_{0(3)}^2 
\left(\alpha_2+\frac{\alpha'_2g_\sigma\sigma}{2}\right) 
\nonumber \\
&+ \frac{1}{2}m^2_\rho\bar{\rho}_{0(3)}^2
+ \frac{1}{2}{\alpha_3}'g_\omega^2 g_\rho^2\omega_0^2\bar{\rho}_{0(3)}^2
+ p_{kin}^{p,n} + \frac{\mu_e^4}{12\pi^2}
\nonumber \\
&+ \frac{1}{3\pi^2}\int_0^{\sqrt{\mu_\mu^2-m^2_\mu}}\frac{dk\,k^4}{(k^2+m_\mu^2)^{1/2 }},
\label{pressure}
\end{align}
with
\begin{equation}
p_{kin}^{p,n}=\frac{\gamma}{6\pi^{2}}\int_{0}^{k_{F_{p,n}}}\frac{k^4}{(k^2+(M^*)^2)^{1/2}}dk
\label{pkin}
\end{equation}
and 
\begin{align}
&\rho = \frac{1}{2}m^2_\sigma\sigma^2 
+ \frac{A}{3}\sigma^3 + \frac{B}{4}\sigma^4 - \frac{1}{2}m^2_\omega\omega_0^2 
- \frac{C}{4}(g_\omega^2\omega_0^2)^2 
\nonumber\\
&+ g_\omega\omega_0\rho
- \frac{1}{2}m^2_\rho\bar{\rho}_{0(3)}^2
+\frac{g_\rho}{2}\bar{\rho}_{0(3)}\rho_3
- \frac{1}{2}\alpha'_3g_\omega^2 g_\rho^2\omega_0^2\bar{\rho}_{0(3)}^2
\nonumber\\
&- g_\sigma g_\omega^2\sigma\omega_0^2
\left(\alpha_1+\frac{\alpha'_1g_\sigma\sigma}{2}\right) 
- g_\sigma g_\rho^2\sigma\bar{\rho}_{0(3)}^2 
\left(\alpha_2+\frac{\alpha'_2 g_\sigma\sigma}{2}\right) 
\nonumber \\
&+ \varepsilon_{kin}^{p,n} + \frac{\mu_e^4}{4\pi^2}
+ \frac{1}{\pi^2}\int_0^{\sqrt{\mu_\mu^2-m^2_\mu}}dk\,k^2(k^2+m_\mu^2)^{1/2},
\label{denerg}
\end{align}
where
\begin{equation}
\varepsilon_{kin}^{p,n}=\frac{\gamma}{2\pi^{2}}\int_{0}^{k_{F_{p,n}}}k^2(k^2+(M^*)^2)^{1/2}dk.
\label{ekin}
\end{equation}
For this work, we use the muon mass $m_\mu = 105.7$~MeV, massless electrons, and momentum $k$. The quantities $p_{kin}^{p,n}$, Eq.~\eqref{pkin}, and $\varepsilon_{kin}^{p,n}$, Eq.~\eqref{ekin}, are the kinetic terms for pressure and energy density, respectively. The indices $p,n$ stand for protons ($p$) and neutrons ($n$),  $k_{F_{p,n}}$ is the Fermi momentum, and $\gamma$ is the degeneracy factor($\gamma = 2$ for asymmetric matter). The effective mass for the nucleon is $M^* = M - g_\sigma \sigma$.

The parameterizations used in this work are considered type $\sigma^3 + \sigma^4 + \omega^4 +$ {\it crossed terms models}~\cite{PRC90-055203-2014}, named as FSUGold2~\cite{fsugold2} and Z271v6~\cite{z271v6}.
 
\section{Numerical Analysis\label{sec:Numerical-Analysis}}

In order to determine the mass-radius relation related to the star, we need to solve the teleparallel equations, namely Eq.~\eqref{eq:ubarprime} and Eq.~\eqref{eq:pbarprime}. For this, it is necessary to consider the charge neutrality and $\beta$-equilibrium conditions. To describe this matter, we consider the existence of protons, neutrons, electrons, and muons. The muon threshold is directly related to the chemical potential of the electron such that $\mu_e = (3\pi^2n_e)^{1/3}> m_\mu$, where $n_e$ is the electron density. From these assumptions, we can write the conditions 
\begin{eqnarray}
\mu_n - \mu_ p + \mu_e &=& 0 \quad \mbox{and} \nonumber \\
n_p - n_e - n_\mu &=& 0,
\label{conditions}
\end{eqnarray}
where $\mu_e = \mu_\mu$ and $n_\mu = \left[(\mu_\mu^2 - m_\mu^2)^{(3/2)}\right]/3\pi^2$. The equations for total pressure and energy density of stellar matter are given by Eq.~\eqref{pressure} and Eq.~\eqref{denerg}, using the conditions in Eq.~\eqref{conditions}. For the solution of those equations, the following conditions were considered: $p(r=0) = p_c$ (central pressure) and $u(r=0) = 0$ in the center of the star; at the surface: $p(r=R) = 0$ and $u(r=R)\equiv M$, where $R$ and $M$ are the radius and mass of the star, respectively. To describe the neutron star crust, the Baym-Pethick-Sutherland equation of state~\citep{bps} for densities between $0.158\times10^{-10} \leqslant\rho\leqslant 0.891\times10^{-2}$~fm$^{-3}$  was used.

As we will see below, we constructed the mass-radius relation for both EoS, comparing them with observational constraints. Furthermore, we studied the behavior of the mass and radius as functions of $\beta_3$.

As stated in Section~\ref{sec:Modified-Teleparallel-Gravity}, the difference between equations of (\ref{eq:Conj_bar}) and their equivalent ones in TEGR is the presence of an effective gravitational coupling constant [cf. Eq. (\ref{eq:rhopeffec})]. Since our model has a free parameter $\beta_3$, it plays a central role in our model. 

\begin{figure}[!htb]
\centering
\includegraphics[width=\columnwidth]{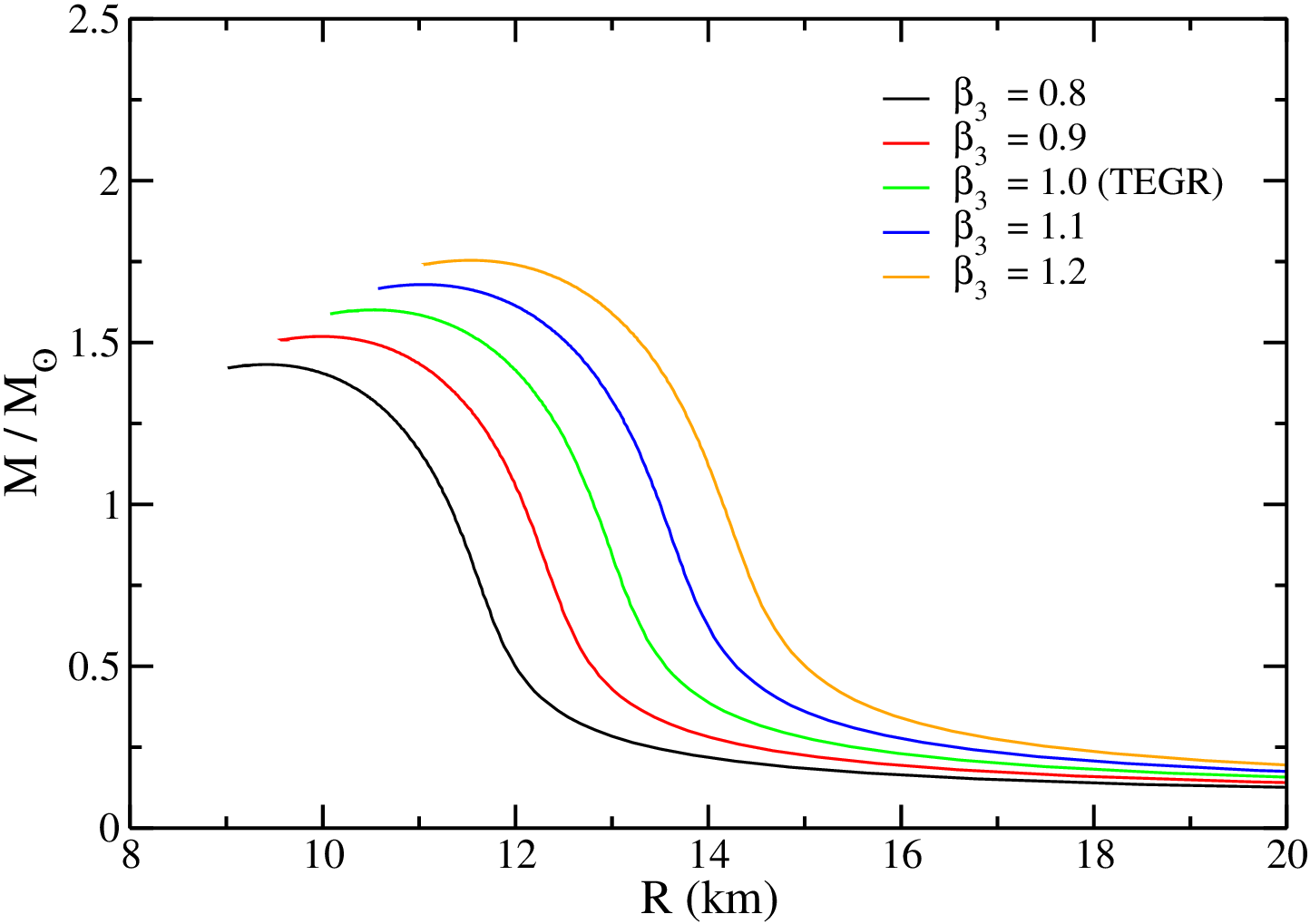}
\caption{Mass-radius diagram for Z271v6 parameterization with different
values of $\beta_3$.}
\label{fig:MxR-z271}
\end{figure}

In this sense, the next two figures show the mass-radius relations for different values of $\beta_3$. In Fig.~\ref{fig:MxR-z271}, five curves for Z271v6 parameterization are plotted. The lower curve with $\beta_3=0.8$ has a maximum value for mass around $1.43M_{\odot}$ and the upper curve plotted with $\beta_3=1.2$ has a value around $1.75M_{\odot}$. 

\begin{figure}[!htb]
\centering
\includegraphics[width=\columnwidth]{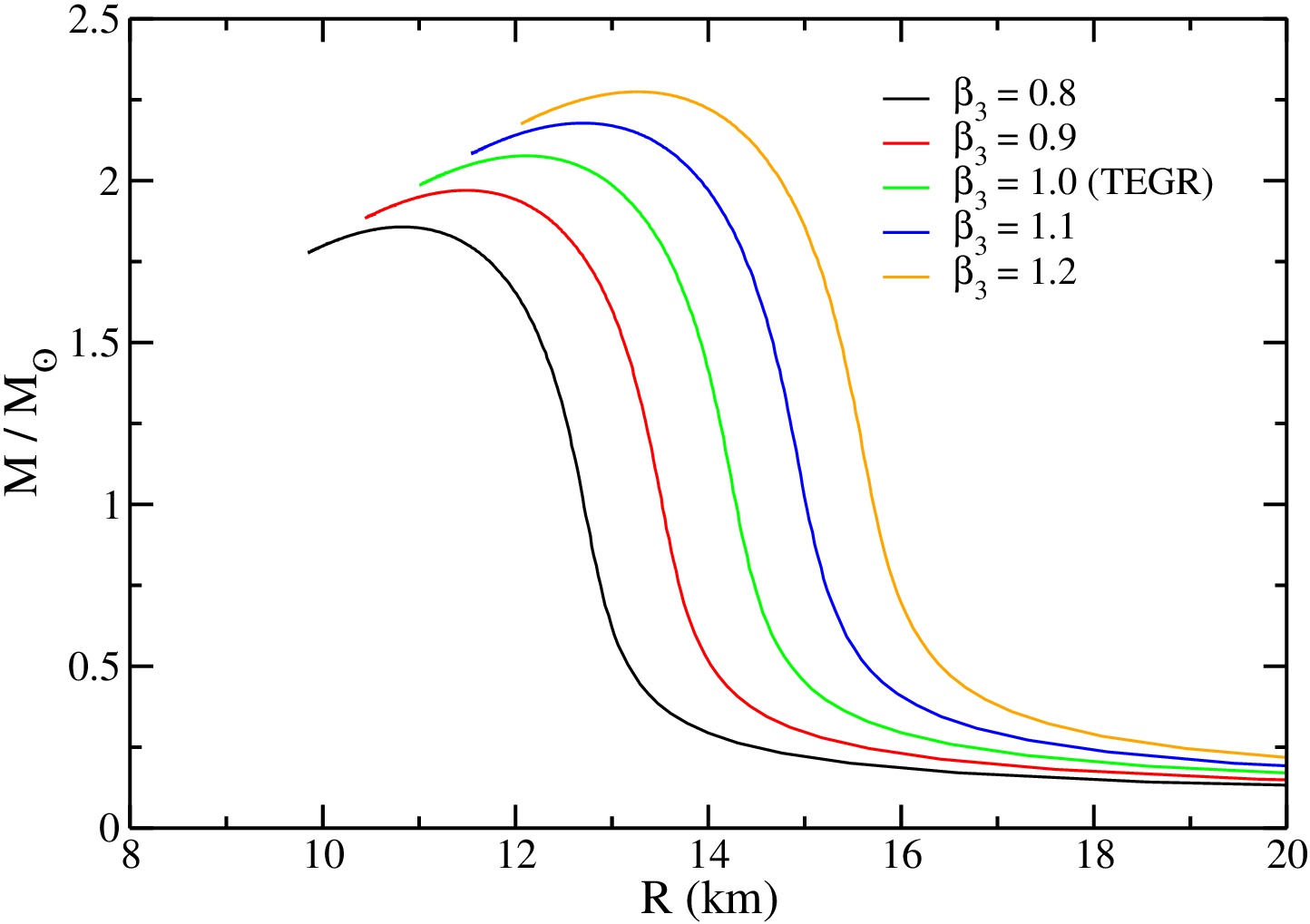}
\caption{Mass-radius diagram for FSUGold2 parameterization with different
values of $\beta_3$.}
\label{fig:MxR-FSU}
\end{figure}

In Fig.~\ref{fig:MxR-FSU} the behavior of the FSUGold2 parameterization is presented. In this plot can be seen five curves in which the lower curve plotted in $\beta_3=0.8$ has a maximum value for mass around $1.85M_{\odot}$ and the upper curve plotted in $\beta_3=1.2$ has a value around $2.27M_{\odot}$. 

As mentioned before, we also can see the effect of the chosen parameterization of the EoS in TEGR for $\beta_3=1$. In this specific case, the results are the same as GR, where the maximum value for mass is $1.60M_{\odot}$ for Z271v6 and $2.08M_{\odot}$ for FSUGold2~\cite{PRC93-025806(2016)}. Note that, for both parameterizations, we have an increase/decrease behavior of the mass (and consequently of the maximum mass) when we increase/decrease the parameter $\beta_3$. The same occurs with the radius of the star.

In order to better understand this behavior, we varied the value of $\beta_3$ and generated the curves for the maximum values obtained in the radius-mass diagram.

\begin{figure}[!htb]
\centering
\includegraphics[width=\columnwidth]{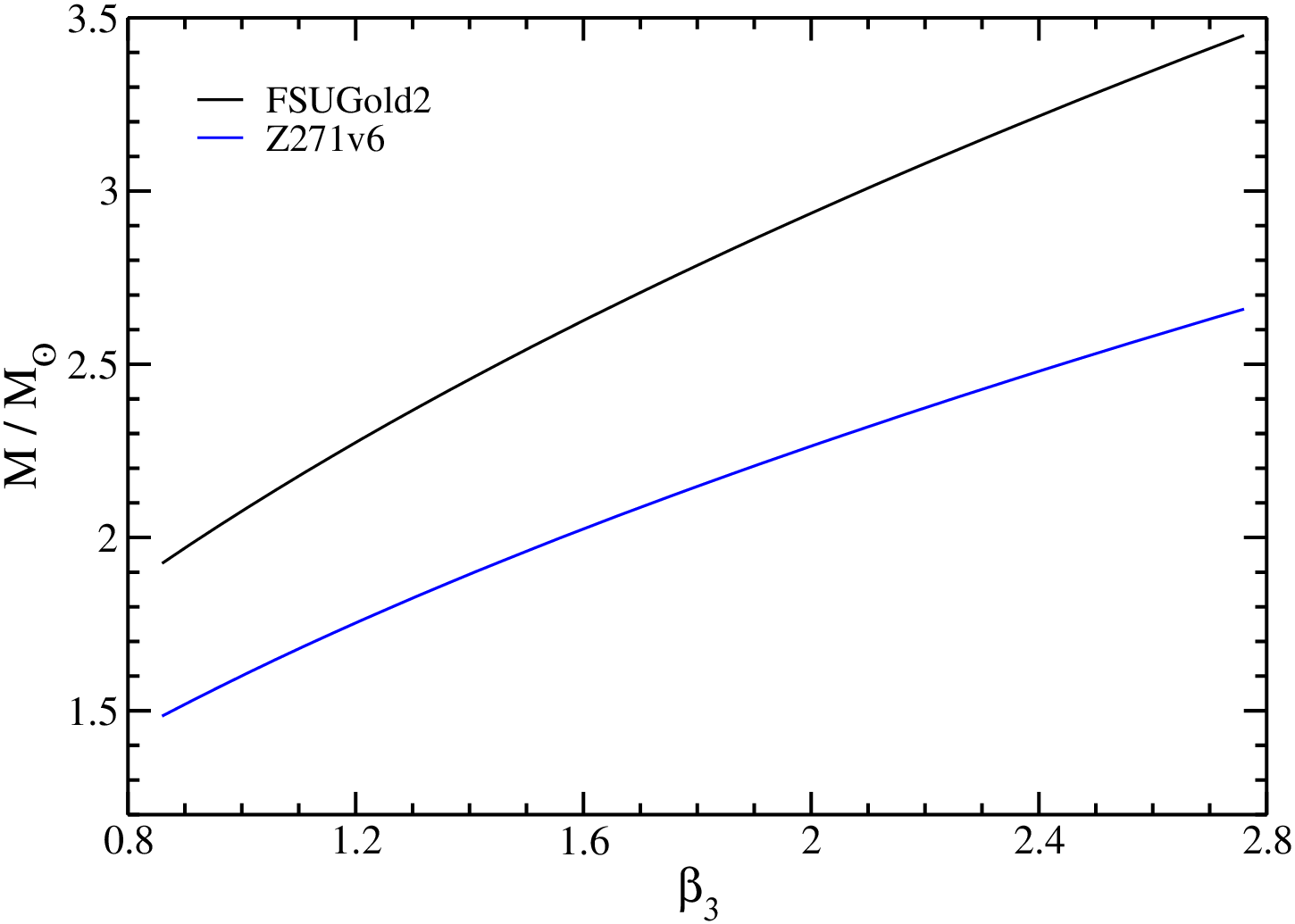}
\caption{Maximum mass as a function of $\beta_3$ for FSUGold2 and Z271v6 parameterizations.}
\label{fig:Mxbeta3}
\end{figure}

In Fig.~\ref{fig:Mxbeta3}, we see how the maximum masses vary in the interval $\beta_3\in [0.86,2.76]$ for both Z271v6 and FSUGold2 parameterizations of the EoS. In both cases, an increase in the value of the parameter $\beta_3$ implies higher values for the maximum masses. In particular, the maximum values obtained with the FSUGold2 parameterization are typical $30\%$ greater than the values obtained with Z271v6.

\begin{figure}[!htb]
\centering
\includegraphics[width=\columnwidth]{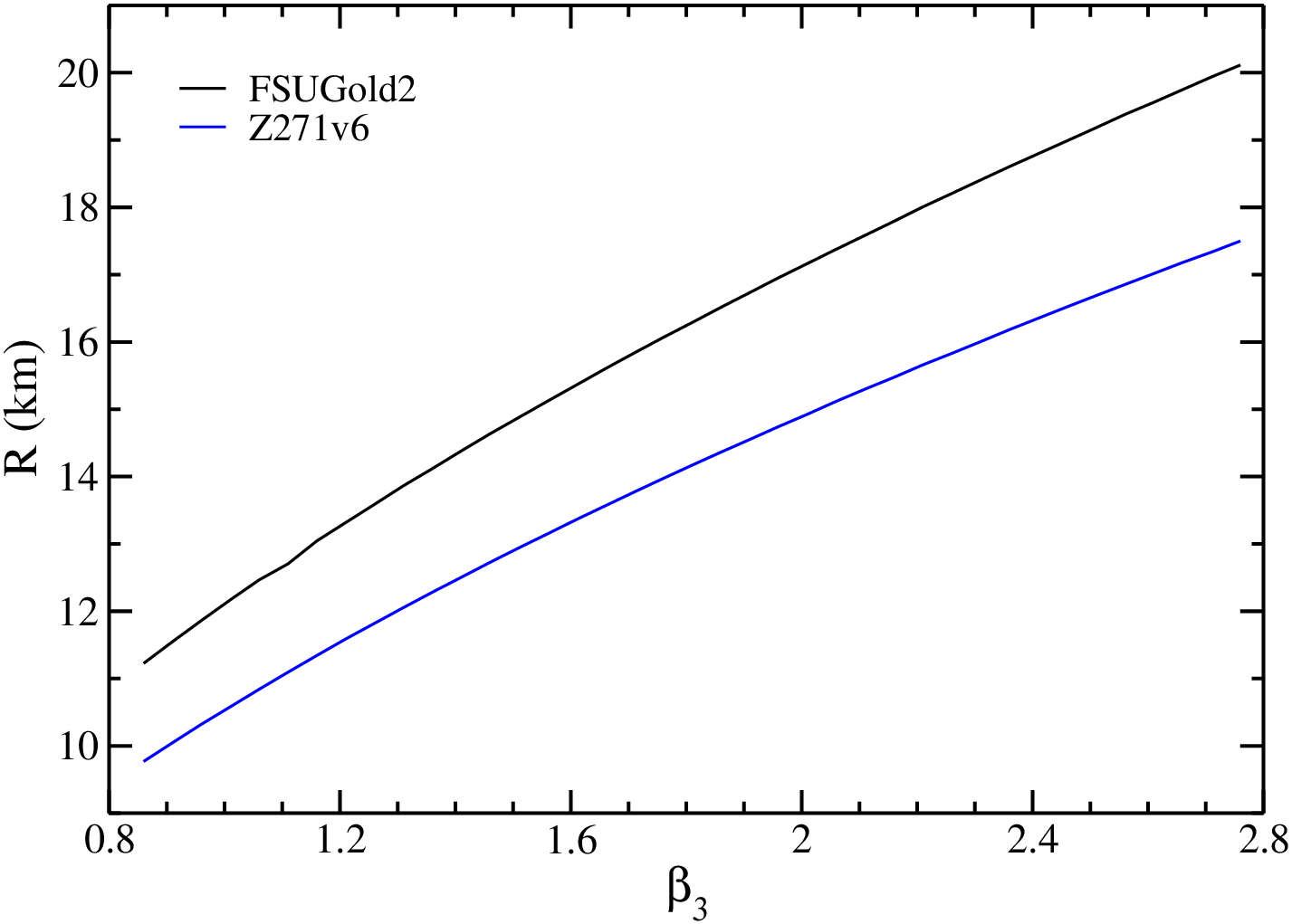}
\caption{Radius as a function of $\beta_3$ for Z271v6 and FSUGold2 parameterizations.}
\label{fig:Rxbeta3}
\end{figure}

The radius and the free parameter are also correlated. As we can see in Fig.\ref{fig:Rxbeta3}, the radius also increases with higher values of $\beta_3$ in both parameterizations of the EoS. The increasing of the radius with $\beta_3$ observed is an expected result, once mass and radius are correlated. The values from $\beta_3=0.86$ to $\beta_3=2.76$ expresses radius from $9.76$~km to $17.49$~km for Z271v6 and from $11.23$~km to $20.11$~km for FSUGold2. The values of radius obtained with FSUGold2 parameterization are roughly $15\%$ greater than those obtained with Z271v6.

For $\beta_3=1$, we recover TEGR, as consequence, the results from GR are encapsulated in our model. Hence, we can also compare both parameterizations in our model with the ones obtained in GR. 

A natural step of our study is to compare our results with observational data~\citep{Abbott(2020),Miller(2021),Riley(2021),Fonseca(2021)}.

\begin{figure}[!htb]
\centering
\includegraphics[width=\columnwidth]{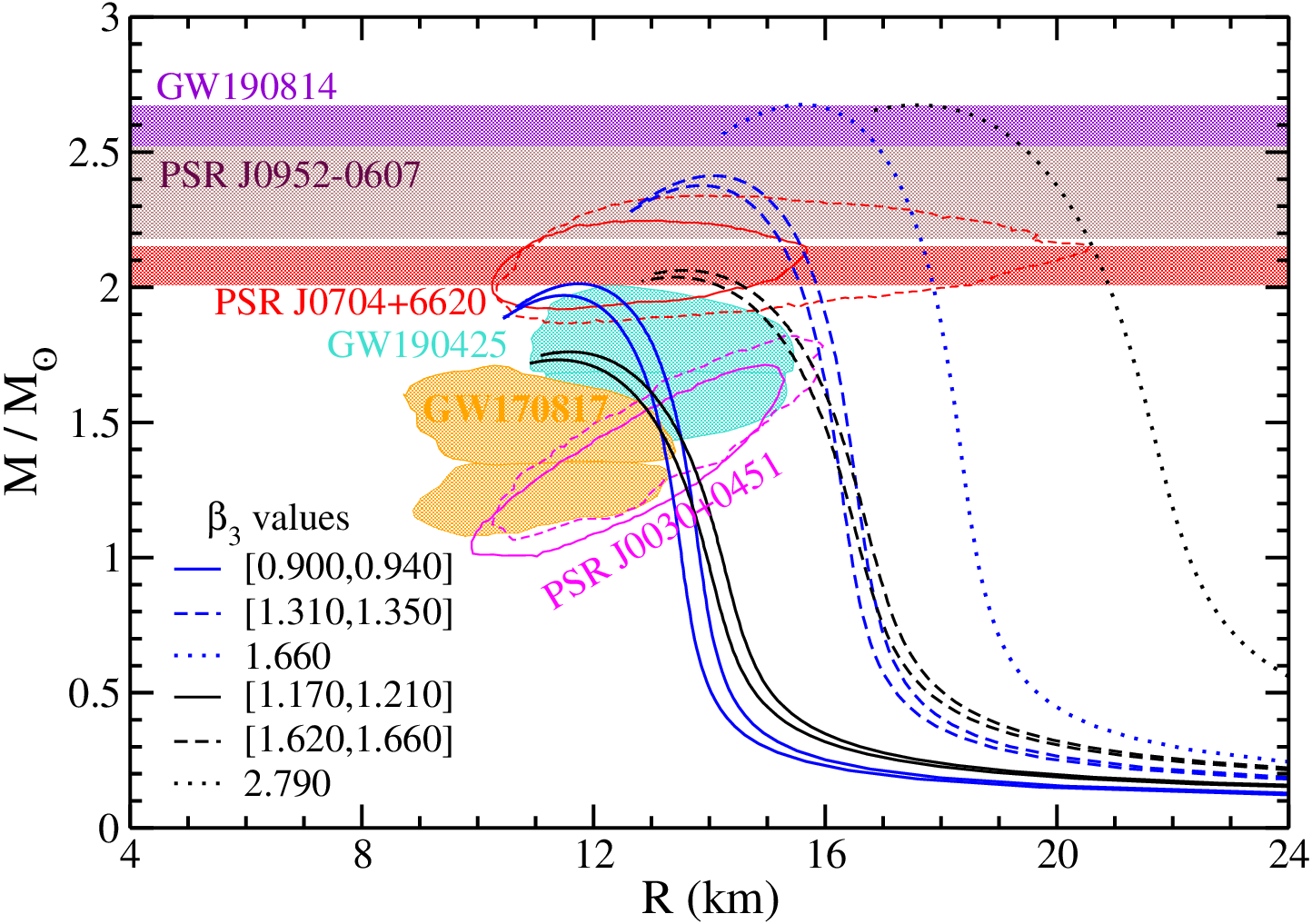}
\caption{Mass-radius diagram constructed for the FSUGold2 in blue, and Z271v6 in black for different values of $\beta_3$. The contours are related to data from the NICER mission, namely, PSR~J0030+0451~\citep{Riley(2019),Miller(2019)} and PSR~J0740+6620~\citep{Riley(2021),Miller(2021)}, the GW170817 event~\citep{Abbott(2017),Abbott(2018)}, and the GW190425 event~\cite{Abbott(2020)-2}, all of them at $90\%$ credible level. The red horizontal lines are related to PSR~J0740+6620~\citep{Fonseca(2021)}. The brown horizontal lines are related to PSR~J0952-0607~\citep{Romani(2022)}. The violet horizontal lines refer to recent observational constraints on neutron star mass GW190814~\citep{Abbott(2020)}.}
\label{fig:MxR-constraints}
\end{figure}

In Fig.~\ref{fig:MxR-constraints}, selected mass-radius curves for different values of $\beta_3$ are presented. The observational data allow us to obtain intervals for this parameter in our modified teleparallel gravity for each parameterization used. We verify a range of the parameter $\beta_{3}$ for each EoS. For FSUGold2, we observe a lower $\beta_{3}$ at 0.43 and higher $\beta_{3}$ at 2.170, where the astrophysical events have your respective intervals
as follows: GW190814 has values among $1.480\leq\beta_{3}\leq1.660$; PSR~J0952-0607, $1.100\leq\beta_{3}\leq1.450$; PSR~J0704+6620, $0.930\leq\beta_{3}\leq1.060$; PSR~J0740+6620 contour, $0.820\leq\beta_{3}\leq2.170$; GW190425, $0.710\leq\beta_{3}\leq1.250$; GW170817, $0.430\leq\beta_{3}\leq0.920$; PSR~J0030+0451, $0.500\leq\beta_{3}\leq1.330$. In Z271 EoS, we observe
higher values for $\beta_{3}$ at each astrophysical event as we can verify in the following: GW190814, $2.480\leq\beta_{3}\leq2.790$; PSR~J0952-0607, $1.850\leq\beta_{3}\leq2.440$; PSR~J0704+6620, $1.570\leq\beta_{3}\leq1.800$; PSR~J0740+6620 contour, $1.360\leq\beta_{3}\leq2.790$; GW190425, $0.590\leq\beta_{3}\leq1.640$; GW170817, $0.580\leq\beta_{3}\leq1.190$;
PSR~J0030+0451, $0.640\leq\beta_{3}\leq1.720$ -- the higher values for $\beta_3$ are necessary to accommodate data from \citep{Abbott(2020)}. As consequence, for Z271v6, the range of values for $\beta_3$ is $0.58\leq \beta_3\leq2.79$ -- the values of $\beta_3$, in this case, are greater than those obtained in FSUGold2.

\section{Final Remarks\label{sec:Final-Remarks}}

In this work, we studied a modified teleparallel gravity for describing compact objects like neutron stars. Our modified model proposed a general linear combination of the quadratic invariant build with the torsion tensor which composes the TEGR Lagrangian. We noted the field equations structure remained the same as TEGR with the replacement of $S^{\mu\nu\rho}\rightarrow\Sigma^{\mu\nu\rho}$ given in Eq. (\ref{eq:NewSigma}).

Analyzing the field equation for static spherically symmetric compact objects, we demonstrated that the conservation equation remains valid only if we constrain our parameters in a specific way [cf. Eq. (\ref{eq:constraint})]. That allowed us to redefine an effective gravitational coupling constant depending only on a free parameter $\beta_3$, {[}Cf. Eq. (\ref{eq:rhopeffec}){]}. Hence, we could rewrite our set of TOV-like equations in a form similar to those found in TEGR.

In the sequence, we studied the behavior of mass and radius as functions of the free parameter $\beta_3$ for two distinct parameterizations of the EoS. In this analysis, it was possible to see that both the values of maximum mass and radius increase with $\beta_3$.

The comparison of our results  with observational data enabled us to establish ranges for our free parameter. The range is different for each parameterization of the EoS. For FSUGold2 EoS, the range is $0.430\leq \beta_3\leq2.170$;  for Z271v6, we obtained $0.580\leq \beta_3\leq2.790$. It is interesting to note that both parameterizations still accommodate TEGR (i.e. $\beta_3=1$).  Also, we verify that there is an interval of values of $\beta_3$ which accommodates results from both FSUGold2 and Z271v6 parameterizations, namely $0.58\leq \beta_3\leq2.17$. In order to distinguish which parameterization  of the EoS is more adequate to describe NS in our model,  we need to apply our equations to other physical systems. A particularly promising area  to do so is cosmology. If an independent range of values for $\beta_3$ is obtained, we should be able to compare it with the ranges obtained in this paper and then reanalyze the role of the EoS parameterizations for describing compact objects.

\section*{Acknowledgments}
S.~G.~V and P~.J.~P. are grateful to Conselho Nacional de Desenvolvimento Cient\'ifico e Tecnol\'ogico (CNPq) for financial support under Grant No. 400879/2019-0. S.~B.~D. also grateful to CNPq for financial support. M.~D. also acknowledges CNPq under Grant No. 308528/2021-2 and Funda\c{c}\~ao de Amparo \`a Pesquisa do Estado de S\~ao Paulo (FAPESP) under Thematic Project 2017/05660-0 and Grant No. 2020/05238-9. This study was financed in part by the Coordena\c c\~ao de Aperfei\c coamento de Pessoal de N\'ivel Superior - Brazil (CAPES) - Finance Code 001 - Project number 88887.687718/2022-00 (M.~D.). This work is a part of the project INCT-FNA proc. No. 464898/2014-5.

\end{document}